\documentclass[12pt]{iopart}
\usepackage{bm}
\begin{document}

\def\bD{{\bf D}}
\def\beps{{\bm \epsilon}}
\def\cst{C^{*}}
\newcommand{\dbar}{d\mkern-6mu\mathchar'26} 

\title{Thermodynamic entropy of a many body energy eigenstate}

\author{J. M. Deutsch}
\address{
Department of Physics, University of California, Santa Cruz, California 95064}


\begin{abstract}
It is argued that a typical many body energy eigenstate has a
well defined thermodynamic entropy and that individual eigenstates
possess thermodynamic characteristics analogous to those of generic
isolated systems. We examine large systems with  eigenstate energies
equivalent to finite temperatures.  When quasi-static evolution of
a system is adiabatic (in the quantum mechanical sense), two coupled
subsystems can transfer heat from one subsystem to another yet remain in an
energy eigenstate.  To explicitly construct the entropy from the wave
function, degrees of freedom are divided into two unequal parts.  It is
argued that the entanglement entropy between these two subsystems  is the
thermodynamic entropy per degree of freedom for the smaller subsystem.
This is done by tracing over the larger subsystem to obtain a density
matrix, and calculating the diagonal and off-diagonal contributions
to the entanglement entropy.
\end{abstract}

\pacs{
}

\maketitle

\section{Introduction}
\label{sec:intro}

The main question investigated here is if it is possible to define a {\em thermodynamic} entropy for {\em energy eigenstates}
of a generic many body system, and if so,  how this can be characterized by examining the wave function in such a state.
The thermodynamic entropy $S$, is measurable experimentally by changing parameters
such as temperature $T$, and calculating small changes in heat $\dbar Q$ via the relation
\begin{equation}
dS = \dbar Q/T .
\end{equation}
By incrementally measuring $dS$ relative to a reference state, say at low temperature, the
entropy can be calculated.
The entropy is important because it is only a function of thermodynamic state variables, not
a system's history, and therefore it should be, in principle, calculable without changing
external parameters. Classically, the Boltzmann hypothesis, that the entropy is
the logarithm of the number of states dynamically accessible to a system, appears to
be correct for generic systems in strict thermal equilibrium~\cite{MaStatMech}. This gives a means
to determine system's entropy with external parameters held constant. 

In quantum mechanics, our understanding of entropy is not as well developed.
A theoretical calculation of the entropy is possible
via the free energy, using the canonical ensemble. An ensemble of systems are
summed over, each one with a different energy. Although this appears to work
in practice, the reason why this {\em canonical entropy} formula works for a pure state is not clear. 
The calculation assumes 
the system is in a mixed state, with microcanonical or canonical weights.
However one expects that thermodynamics should apply to pure states, so
it is far from clear that the canonical entropy calculation applies to that case.
The Von Neumann entropy for a system with a density matrix $\rho$ is 
$S_{VN}(\rho) = - \tr \rho \ln \rho$,
and is {\em zero} for a system in a pure state, although it gives the usual result
for the entropy using a canonical mixed state. As a result, $S_{VN}$ is
clearly not a candidate for the thermodynamic entropy of a pure state.

On the other hand, if we consider a system in an energy eigenstate, it has a trivial
time dependence, and it is hard to see how to associate a thermodynamic entropy
to it, in any way similar to the Boltzmann hypothesis. There is no exploring of phase space with time
and it is not clear from our intuition of heat, how the concept of 
entropy could be meaningful.

In this work I consider a single energy eigenstate of a large ergodic system, with
the word ``ergodic" defined in the next section. I argue that for such a state,
the concept of a thermodynamic entropy is still meaningful and that
such a system under extremely slow external perturbations such as
a changing magnetic field, will transfer heat between different parts
of a system, while remaining in an energy eigenstate in accordance with
the usual laws of thermodynamics. 

Because this entropy is a function of the state of the system, it should
be possible to find a recipe to calculate its thermodynamic entropy from
the wave function itself with no changes in external parameters. This
recipe should be equal to the entropy calculated in the canonical ensemble
for large systems.  This lends support to the conjecture that even an
energy eigenstate has thermodynamic properties seen for more generic
density matrices.

To construct the entropy of an energy eigenstate, the quantity used is the  ``entanglement entropy". First 
we subdivide a system into two macroscopic
systems, one labeled $\theta$ and the other labeled $\phi$  with $m$ and $n$ degrees of freedom respectively. 
One firsts defines a density matrix $\rho_\phi = \Tr_\theta |\phi,\theta\rangle\langle\phi,\theta|.$ 
Then the entanglement entropy $S_E \equiv  -\tr \rho_\phi \ln \rho_\phi$. 
This has many useful properties. For example it is also equal to 
$ - \tr \rho_\theta \ln \rho_\theta$. 

The entanglement entropy plays a crucial role in quantum information theory~\cite{BennetQuantInfo} and
has also been used in the study of black holes~\cite{Bekenstein}, and quantum phase transitions~\cite{Vidal}. 
The entanglement entropy has been argued to be a unique measure of entanglement in
a pure state~\cite{PopescuRohrlich} by making analogies with thermodynamics, but
it's relationship to the thermodynamic entropy is still unclear.

At zero temperature, $S_E$ is non-zero but non-extensive
and for a variety of systems is a power law of system
size~\cite{SrednickiEntanglement,WolfeEntanglement}.  
At finite temperature and large system size,
1+1 dimensional quantum field theories give extensive behavior.

Here we examine examine how to calculate the entanglement entropy
in an energy eigenstate by means similar to those used earlier to
study the question of energy eigenstate thermalization. First in
Sec. \ref{sec:EnEigTherm} we discuss this problem.  Applying that
conjecture, we show in \ref{sec:WaveFunctionThermo}  how this implies
that the notion of heat and thermodynamic entropy are meaningful for
energy eigenstates. Finally in Sec. \ref{sec:AnalEntEntr} we present a
calculation of the entanglement entropy.

We argue that for a system in an energy eigenstate where the number of
degrees of freedom go to infinity, and energies corresponding to a fixed
finite temperature, the entanglement entropy becomes the thermodynamic
entropy per degree of freedom. This gives an explicit formula for deducing 
thermodynamic properties from individual energy eigenstates.

The outline of this paper is as follows. In Sec. \ref{sec:EnEigTherm}, we
discuss previous work that argues that for energy eigenstates, the expectation value of
a large class of observables will be equivalent to averages
using the microcanonical ensemble. Using this, in Sec. \ref{sec:WaveFunctionThermo} 
we explore the relation between thermodynamic entropy and its microcanonical value,
for systems in energy eigenstates. In  Sec. \ref{sec:AnalEntEntr} we analyze
the entanglement entropy, using methods similar to those of Sec. \ref{sec:EnEigTherm}.

\section{Energy Eigenstate Thermalization}
\label{sec:EnEigTherm}

Previous work by the author~\cite{DeutschQuantStatMech} and others~\cite{SrednickiQuantStatMech}
attempted to understand why the laws of quantum statistical mechanics work for
an isolated system.  The approach
taken was to see what could be derived about statistical mechanics
from quantum mechanics without any additional assumptions and for a particular
choice of model systems.
The assumption made in quantum statistical mechanics is that the
average over time $\langle \dots \rangle_t$ of some observable quantity 
$ \langle \psi | A | \psi  \rangle $ is equal to a microcanonical average at 
a total energy $e$ that is assumed to be well defined (see below),
\begin{equation}
\label{eq:psiApsiMicro}
\langle\langle \psi | A | \psi  \rangle\rangle_t = \sum_j \Delta(e,e_j ) \langle j|A|j\rangle
\end{equation}
where $e_j$ labels an energy eigenstate of the entire system and 
$\Delta (e,e_j ) $ is a function that is sharply peaked at $e = e_j$.  
For a system containing a large number of degrees of freedom and for
a large class of operators $A$, this can be 
written with negligible error in terms of the canonical distribution
at fixed temperature (throughout this work, units are chosen so that Boltzmann's constant is unity.)
\begin{equation}
\label{eq:psiApsiCan}
\langle\langle \psi | A | \psi \rangle\rangle_t = \frac{\sum_j e^{- e_j/T}\langle j|A|j\rangle}{\sum_j e^{- e_j/T}}
\end{equation}
We will postpone to the end of this section how these formulas
should, rather simply, be modified
to take into account fluctuations in the total energy, but Eq. \ref{eq:psiApsiMicro}
or equivalently Eq. \ref{eq:psiApsiCan} has been enormously successful in explaining problems in almost
every branch of physics. For the purposes of this paper, systems obeying 
this equation will be called ``ergodic''. 

In classical mechanics, a system with a few of degrees of freedom
such as Sinai billiards, have time averages given by the microcanonical
distribution.
A quantum mechanical treatment of the same system 
cannot be expected to give the microcanonical
distribution. It is easy to show by counterexample, that one needs at least one
more requirement; the number of degrees of the system must also be large.
Indeed, if the spacing between energy levels is not small it is impossible
to define a microcanonical distribution in a precise way. 

Indeed, the density of states $G(E)$ for a system with $n$ degrees of
freedom is related to the entropy, and for an extensive system $G(E)
 = (1/e_0) \exp(n s(E/n))$. Here $s$ is the entropy per particle and $e_0$ is
an energy normalization. Therefore,
if the energy width in the microcanonical average is $\delta$, then the
number of states being averaged over is proportional to $\delta \exp(n s(E/n))$. 
With fixed $\delta$, the number of states contributing to the
average diverges exponentially with $n$ implying that fluctuations in
microcanonical quantities will rapidly go to zero with increasing $n$
as the total energy is varied.

Having a large number of degrees of freedom however, is not enough to ensure 
ergodicity. As a simple example, consider a perfect harmonic crystal in which case the
initial choice of wave function alters the time averages of an observable. If such
a system were to start with a wave function obeying Eq. \ref{eq:psiApsiMicro},
then shining light on it so as to couple to some modes preferentially, will now 
violate this microcanonical average.

Thus we must search for a mechanism that can explain how an experimental system can continue to
give microcanonical averages for time averaged quantities despite atypical initial
states, such as the example of light described above.
The approach taken was similar to understanding how this works for an almost
ideal classical gas.  A genuine ideal gas has no interaction between different
particles and therefore will not be ergodic, but can be made so 
by slight modification. For example, the
particles can be given hard cores of very small diameter,  which will have a
negligible effect on the statistical and thermodynamic properties computed from
the Gibbs distribution.  However after a long enough time, the system will
explore almost all of its available phase space enabling the  rigorous application of
the statistical mechanical formula Eq. \ref{eq:psiApsiMicro}.  In the same spirit, suppose we start with a Hamiltonian that
decouples into n separate subsystems
\begin{equation}
H_ 0 =~ \sum_{i=1}^n h_0 (x_i , p_i )
\end{equation}
This system is not ergodic. However the main result of the previous
work~\cite{DeutschQuantStatMech}, was to show that with negligible error
in the limit of large n, it can be made ergodic by the addition of a
small perturbation as described below.  

The model considered was 
\begin{equation}
H = H_0 + H_1
\end{equation}
where $H_1$ is added in the hopes of
making the system ergodic. In the case of an ideal gas for example, one may
want to add some interaction between the different particles, for example, take
\begin{equation}
H_1 = \sum_{i<j}^n V(r_i - r_j)
\end{equation}
However this is extremely hard to analyze so
instead of adding in these interactions explicitly, we use a 
random matrix model to understand the effects of a generic perturbation.
If we consider the problem in the basis of energy eigenvalues of $H_0$,
then we model $H_1$ by a
real symmetric matrix whose elements are chosen from a  real random Gaussian
ensemble, with certain physically sensible conditions
on the magnitude of the elements described in the next paragraph. The use
of a random matrix is sensible in this context as much work, starting
with the monumental work of Wigner~\cite{WignerReview}  that shows a deep connection between the physics of
interacting systems and random matrices. 

We take the variances of these random couplings to decrease 
away from the matrix diagonal.
\begin{equation}
h_{ij} \equiv \langle E_i |H_1 | E_j\rangle ~,~\overline {h_{ij} h_{kl}} = \epsilon_{i-j}^2 \delta_{ik} \delta_{jl} ,
\end{equation}
where the magnitudes of the $\epsilon$ are taken to be much less than $T$ but much greater than the energy spacing as will be
discussed in more detail below. In this paper, the line above the matrix elements denotes an average over the ensemble
of random matrices.

The reason that the variance of the matrix elements are taken to depend on position
is explained in more detail in Appendix \ref{app:RangeOfPotential}, but in general, one expects 
asymptotically that the effect of coupling from states of different 
energies $E_1$ and $E_2$,  to decrease with their energy difference. The size of elements are diminished by a phase space factor
\begin{equation}
\langle E_1 | H_1 | E_2 \rangle \sim  e^ { | E_1 - E_2 |  / T }
\end{equation}
for $ T \ll | E_1 - E_2 | \ll E_1 $ (the ground state energy of the
system is set to 0).  The temperature is defined by the usual prescription
\begin{equation}
\frac{1}{T} = \frac{ \partial S } { \partial E }
\end{equation}
The energy $E$ in the above derivative can be evaluated at either 
$E_1$ or $E_2$ since for a
large system $T \ll E$.  So when $ | E_1 - E_2 |  >> T $, the proportion
of non-zero matrix elements is effectively zero. To simplify the
model further, we consider
a banded random matrix, where the width of the band increases with energy. Inside
the band, all off-diagonal elements have the same variance $\epsilon$.
The precise form of the cutoff is unimportant to the conclusions but should
be present on physical grounds and also prevents unphysical divergences in 
expectation values. 

The random matrix model defined above is similar to one analyzed by
Wigner~\cite{WignerBanded}, where he considered a matrix with diagonal elements
that were linearly increasing, $D_i = \Delta i$, and $\Delta = 1/G(E)$ can be taken to
be the average energy level spacing,  and off-diagonal random matrix elements 
are banded as above with the width $n_b \gg 1$.
Eigenvectors are random but the square of the amplitudes are well
defined and are not constant. Denoting the amplitudes of the $ith$ eigenvector
by $c_{ij}$, 
\begin{equation}
\label{eq:sigsq}
    \sigma_{ij} \equiv \overline{ |c_{ij}|^2 } = \frac{\epsilon^2}{(\Delta i - \Delta j)^2 + \delta^2}
\end{equation}
where
\begin{equation}
\label{eq:delta=epssq}
	\delta = \frac{\pi \epsilon^2}{\Delta} ,
\end{equation}
for $|i-j|\Delta \ll n_b \Delta$. In the opposite limit, the eigenvalues decay faster
than an exponential. This is why finite $n_b$ prevents unphysical divergences.

As it turns out, the width of the Lorentzian $\delta$, determines the energy width 
that is used when doing a microcanonical average in Eq. \ref{eq:psiApsiMicro}.
Therefore one would like $\delta \ll T$, so as not to change the microcanonical result.
But conversely, $\delta \gg \Delta$ (the energy spacing). 
Both requirements are easily satisfied because as mentioned above, the total number
of states contributing to the microcanonical average is $\sim \delta\exp(n s(E/n))$
This means one should choose $  \Delta = e_0 \exp(-n s(E/n)) \ll \delta \ll T$ which is easily
satisfied for large $n$, as we will take $\delta$ to be independent of $n$. 
In terms of the parameter $\epsilon$ in this model, Eq. \ref{eq:delta=epssq}
gives that $\Delta \ll \epsilon \ll \sqrt{T\Delta}$. Because of the immense smallness of
$\Delta$ the addition of $\epsilon$ will not significantly change the partition
function for this system. Instead it was shown~\cite{DeutschQuantStatMech} 
that its effect is to make infinite time averages in accord with
the microcanonical distribution.

The sense that this model gives microcanonical results is as follows. 
One considers time averages done with one choice of random matrix. The
answer will differ from that of another realization. We can compute what the
variance of the average will be, averaged over all matrices in the ensemble.
It was shown that this variance is proportional to $\Delta/\delta$ which is exponentially 
small in $n$.

The most striking feature of this is that the equivalence to the microcanonical
distribution should apply even to energy eigenstates for a large class of operators $A$. This is surprising because
the time dependence of such wave functions is trivial and therefore does not show
any chaotic time dependent behavior. Of course the spatial dependence is extremely
complex and this is the reason why it can give rise to this kind of self
averaging. This equivalence for energy eigenstates has been recently confirmed
by {\em ab initio} numerical tests~\cite{RigolDunjkoOlshanii} and is often
referred to as ``Eigenstate Thermalization".

A caveat must be stated to the above claim. For an initial state with
a large spread in total energies, the microcanonical distribution is not obtained
because the system cannot be averaged at only one energy. Instead it can be
shown~\cite{DeutschQuantStatMech} that time averages require an additional averaging, over the probability
of finding the system at a particular energy.

\section{Thermodynamics of Energy Eigenstates}
\label{sec:WaveFunctionThermo} 

An energy eigenstate has trivial time dependence and it is of interest to investigate
whether a complete system when placed in such a state still obey thermodynamics. For
an isolated system, there is no heat flow into or out of the system, so we
examine a closed system composed of two macroscopic parts, $A$ and $B$ that are weakly
interacting so that the Hamiltonian is $H = H_A + H_B + H_i$. To analyze this, we will
assume the thermalization of individual eigenstates~\cite{DeutschQuantStatMech},
as argued in the last section. 

Suppose only $H_A$ depends on an external parameter, $x$. For example, $x$ could be
an external magnetic field or the position of a piston. 
Because $H_B$ does not depend on the external parameter,
one could regard $B$ as a heat bath for subsystem $A$, though this would
only be a good analogy when subsystem $B$ was much larger then $A$. For
example, $A$ could be a gas cylinder with a movable piston at position
$x$, and $B$ could be a heat bath in contact with $A$. We can
analyze how energy gets transfered between $A$ and $B$ as a result of
changing $x$ infinitesimally and quasistatically from $x$ to $x+dx$. By
quasistatic, we mean adiabatic in the quantum mechanical sense. We are also
considering a completely isolated system, so that it is adiabatic
in the thermodynamic sense. However the subsystems can transfer energy
between each other.

Because we are only interested in systems that are not integrable (and in fact ``ergodic" in
the sense of the word given above), we expect
that energy level repulsion will prevent any level crossing during a
change in $x$, so it is possible, in principle, to vary a parameter and
stay in an energy eigenstate.

Since in an energy eigenstate $E$  depends on $x$, by the Guttinger-Feynman-Hellman theorem~\cite{FeynmannHellman}
\begin{equation}
\frac{\partial E}{\partial x} = \langle \psi | \frac{\partial H}{\partial x} | \psi \rangle .
\end{equation}
Because of our assumption of individual eigenstate thermalization, 
\begin{equation}
\label{eq:dE=dHmicro}
\frac{\partial E}{\partial x} = \langle\frac{\partial H}{\partial x}\rangle_{m} = \langle\frac{\partial H_A}{\partial x}\rangle_{m}
\end{equation}
where the subscript $m$ under the averages denote a microcanonical average.
The right hand side implies that the energy derivative is the same as for a generic
mixed state (with the usual assumption of a sharply peaked energy distribution). 
Because for a large system the microcanonical average is equivalent to a canonical average,
and the average only involves subsystem $A$, we take the average over only subsystem $A$ using
say, a microcanonical ensemble for $A$. 
Therefore it is possible to relate this to the microcanonical (of canonical) definition
of entropy in $A$ similar to standard procedures, see for example Reif~\cite{Reif}.

First, the work $\dbar W$ done by the system when $x$ is changed quasistatically to $x+dx$ is
\begin{equation}
\dbar W = -\frac{\partial E}{\partial x} dx
\end{equation}
Note that the rate must be slow enough for the system to remain almost entirely in an
energy eigenstate. The work done, say for example by moving a piston, involves
a change in energy of both parts $A$ and $B$. Although the piston is part of
system $A$, a flow of energy from $B$ to $A$  can also occur, and such a flow will
contribute to the total amount of work done in changing $x$. This is familiar in
the common thermodynamics example of the expansion of a gas under adiabatic
conditions, or under constant temperature conditions. In the latter case, heat
flows from a heat bath into the gas contributing to the work done. Therefore the
work done involves the total change in the energy of $A$ and $B$ although the
piston is only attached to $A$.

Second, statistical mechanical entropy of system $A$, $S_{A}$ is defined as $S_{A} = \ln \Omega_A$, where $\Omega_A$ is the number of
states in the energy window being considered. Then as shown by Reif~\cite{Reif}
\begin{equation}
\label{eq:SmicroDeriv}
\frac{\partial S_{A}}{\partial x} = -\frac{1}{T} \langle\frac{\partial H_A}{\partial x}\rangle_{A}
\end{equation}
The microcanonical average is being taken at an energy $E_A = \langle \psi | H_A | \psi \rangle$ for only subsystem $A$.
Appendix \ref{app:RelEntrForce} gives a simple derivation of this.

Then one writes the differential
\begin{equation}
dS_{A} = \frac{\partial S_{A}}{\partial E_A} dE_A + \frac{\partial S_{A}}{\partial x} dx
\end{equation}
Note from Eq. \ref{eq:dE=dHmicro} and \ref{eq:SmicroDeriv} 
\begin{equation}
dS_{A} = \frac{1}{T}(dE_A - \frac{\partial E}{\partial x} dx) 
\end{equation}
The last equality involves $dE_A  + \dbar W$ which is usual definition of the total heat $\dbar Q$ absorbed by $A$.
Hence
\begin{equation}
\label{eq:dSQT}
dS_{A} = \frac{\dbar Q}{T}
\end{equation}

Therefore the ensemble definition of the entropy,
is related to the flow of energy $\dbar Q$ between sub-systems $A$ and $B$ by the usual thermodynamic relation,
Eq. \ref{eq:dSQT}. For thermodynamic purposes, the energy flow
$\dbar Q$ is completely equivalent to heat yet it is seen for a system
in an energy eigenstate. A change in an external parameter
acting on one part of a system in an energy eigenstate will cause entropy to redistribute itself
across subsystems.

Although the above does not show that there is an entropy associated with the
complete system in an energy eigenstate, it shows that entropy changes 
of sub-systems can be induced by a slowly varying external parameter.
The interaction between the sub-systems could also be slowly switched off,
leading to two systems both in energy eigenstates. By first slowly changing
$x$ and then slowly switching off their interaction, one can compare the
difference in entropies for different final values of $x$. This is similar to
the procedure used in determining entropy using reference states. Therefore relative
entropies between different energy eigenstates can be calculated.
This suggests that the absolute entropy for a system in an energy eigenstate
is a meaningful concept. 

In the next section, we argue that this thermodynamic entropy associated with an
energy eigenstate and be calculated by means of the entanglement entropy.

\section{Analysis of the Entanglement Entropy}
\label{sec:AnalEntEntr}

\subsection{The Model}

As discussed earlier in Sec. \ref{sec:intro}, we consider a system in an eigenstate energy $E$ and
with $n_{tot}$ degrees of freedom. We subdivide it into two macroscopic
systems, one $\phi$ and the other $\theta$ with $n$ and $m$ degrees of freedom respectively. We will consider $n$ large
but $m \gg n$. If both sub-systems were uncoupled, we can diagonalize each of them into
energy eigenstates forming a complete set of $|\phi_i\rangle$ and
$|\theta_j\rangle$. The wave function for the complete system can then be written as
\begin{equation}
|\psi\rangle = \sum_{i,j} C_{ij} |\phi_i\rangle|\theta_j\rangle
\end{equation}
The summation is over all states. 

Consider the case where the coupling Hamiltonian  $H_1$ between $m$ and $n$ is weak. Then to first
order in perturbation theory an energy eigenstate of the uncoupled system $|\phi_0\rangle|\theta_0\rangle$ is altered
by $H_1$ as follows
\begin{equation}
\label{eq:1storderpert}
|\psi\rangle = \sum_{i,j} \frac{\langle\theta_j|\langle \phi_i| H_1 | \phi_0\rangle|\theta_0\rangle}{E_0-E_{i,j}} |\phi_i\rangle|\theta_j\rangle .
\end{equation}
As we argued above, $\langle\theta_j|\langle \phi_i| H_1 | \phi_0\rangle|\theta_0\rangle$ will become exponentially small
when the energy difference between the bra and ket states is much greater than $T$. Higher order terms in the perturbation series
also have this property.  Therefore applying a perturbation
that couples $n$ and $m$ will only allow a coupling if their energies differ by 
a microscopic energy of order $T$. 

Because coefficients $C_{ij}$ are identical to those in Eq. \ref{eq:1storderpert}, the only
ones that need be considered are those where the energy of state $|\phi_i\rangle|\theta_j\rangle$
is almost constant.

Although I expect the arguments here to hold more generally, one can make
the model more precise by adopting an approach similar to that used in the
Sec. \ref{sec:WaveFunctionThermo}. We choose the energy eigenvectors as a
basis for $|\psi\rangle$ before coupling $H_1$ is turned on.  Therefore
the energy in an eigenstate $|\phi\rangle|\theta\rangle$  is the sum of
the energy of $\phi$ plus the energy of $\theta$.  When the interaction
$H_1$ is switched on then the effects of this coupling can be modeled
as was done in Sec. \ref{sec:WaveFunctionThermo} as
a banded random matrix.  In this case, $\overline{C_{ij} C_{kl}}$,
averaged over different realizations of $H_1$, are zero for $i \ne k$
or $j \ne l$.

The argument that we present now is analogous to the standard argument of energy exchange between
two subsystems~\cite{MaStatMech}.
Out of states $|\phi_i\rangle|\theta_j\rangle$ that will contribute to $|\psi\rangle$,
there will be ones where the energy $e_\theta$ of subsystem $\theta$ is high and $e_\phi$ is low, and
vice versa. Because the density of states of each subsystem increases extremely
rapidly with energy, there will be a very sharp peak in the number of states
that contribute as a function of $e_\theta$. This means that the two subsystems will
be at the same ``temperature" and the standard deviation in energy of one subsystem $\Delta E$ is $\propto \sqrt{C n m/(n+m)} T$ where
$C$ is the specific heat per degree of freedom. This is much greater than the microscopic thermal energy $T$
but much less than the total energy of $|\phi\rangle$.

\subsection{Entanglement Entropy Calculation}

The density matrix of the $\phi$, in the $|\phi\rangle |\theta\rangle$ basis, tracing over $\theta\rangle$ can be represented
as
\begin{equation}
\label{eq:rhoCoeffs}
\rho_{ij} = \sum_{k} C_{ik} \cst_{jk}
\end{equation}

First we will examine the diagonal portion of this density matrix, $\rho_{ii}$. 
By completeness, $\tr(\rho) = \langle \psi |\psi\rangle = 1$. 
By integrating over a subsystem $m$, $\rho_{ii}$ will be close to zero outside of a
window that depends on the size of $m$. To estimate the size of this window,
we note that total energy of the isolated system is conserved. 
First consider $\rho_{ii}$ averaged over an ensemble of $C's$, (or random matrices) 
\begin{equation}
\label{eq:averhoii}
\overline{\rho_{ii}} = \sum_{j=1}^M \overline{|C_{ij}|^2}
\end{equation}
The number of states effectively contributing, $M$, is finite, because as we just argued, conservation of energy and the very sharp peak
in the density of states as a function of $e_\theta$, imply that only states of $\theta$ with an energy within a few
windows of $\Delta E$, will contribute to this sum. Therefore $M \propto \Delta E \exp(ms(E/n_{tot}))$, $s$ being the
microcanonical entropy per degree of freedom.
In Eq. \ref{eq:averhoii}, we expect that $\overline{|C_{ij}|^2}$ to be a smooth function of the $j$'s in
analogy to Eq. \ref{eq:sigsq}.  Because $\sum_i \rho_{ii} = 1$,
we can estimate the entropy as follows. The number of terms contributing to this sum  is
$N = \Delta E G(E)$ where $G(E)$ is  the density of states of subsystem $\phi$ and is $\propto \exp(n s(E/n_{tot})$.
We take the terms in this window to be constant
so that $\overline{\rho_{ii}} = 1/N$. To get
our initial estimate, we will assume that all off-diagonal components are negligible (but we will
do a better job below.) Therefore this
entropy estimate is then $-\sum_i \rho_{ii} \ln \rho_{ii}$, which is then $ N (1/N) \ln(N) $ 
But $N \propto \sqrt{n} \exp(n s(E/n_{tot}))$. For large $N$, this becomes $n S(E/n_{tot}) + O(\ln n)$. For large
$n$, this becomes precisely the entropy of the subsystem $\phi$. 

The reason why such a crude approximation to the density matrix gives the correct
answer is the same reason as it works with many statistical mechanical calculations. Only the
peak value of $\rho_{ii}$ matters after taking a logarithm. Therefore one could instead
have taken a non-flat distribution for $\rho_{ii}$ and this would not have altered
the dominant term. However by pre-averaging $\rho$, we ignored fluctuations which will
be sizable if $m$ is small. To estimate these we expand $\rho_{ii} = \overline{\rho_{ii}}  + \delta\rho_{ii}$.
Then we average to find the size of the fluctuations. Expanding $\rho_{ii} \ln \rho_{ii}$ in $\delta \rho_{ii}$ and
averaging, the first order terms vanishes, leaving a correction ${\delta \rho_{ii}}^2/(2\overline{\rho_{ii}})$.

With uncorrelated $C_{ij}$'s, $\overline{{\delta \rho_{ii}}^2} = var(\rho_{ii}) \sim M var(|C_{ij}|^2)$. We also expect
that the $C_{ij}$'s will be close to Gaussian, which means that 
$var(|C_{ij}|^2) \propto \langle |C_{ij}|^2 \rangle^2$. Because $\overline{\rho_{ii}} \sim 1/N$, 
$\langle |C_{ij}|^2\rangle \propto 1/(NM)$ which implies $\overline{{\delta\rho_{ii}}^2} \propto M var(|C_{ij}|^2) $. 
Therefore 
\begin{equation}
\sum_i {\delta \rho_{ii}}^2/(2\overline{\rho_{ii}}) =  N \frac{M (\frac{1}{N M})^2}{1/N} = 1/M
\end{equation}
Therefore we expect corrections to the diagonal elements of the density matrix will be exponentially
small in the system size $m$.

We will now turn to a calculation of the contribution of the entanglement entropy due to
off-diagonal matrix elements.
We will first estimate the size of the off-diagonal portions of the density
matrix,  and to do this we will again assume as with a random matrix model, that 
as above, the $C_{ij}$ have random phase and be uncorrelated. 
Averaging over random realizations of the $C$'s gives and estimate for the variance of $\rho_{ij}$
For $i \ne j$, 
\begin{equation}
\label{eq:rhoijEstimate}
\overline{|\rho_{ij}|^2} = \sum_{m,n} \overline{C_{im} \cst_{jm}  \cst_{in} C_{jn}}
 = \sum_{n} \overline{|C_{in}|^2 |C_{jn}|^2} \sim O(N(\frac{1}{NM})^2) = O(\frac{1}{N^2 M})
\end{equation}
Because this is much smaller than the order of $\rho \sim O(1/N)$, it makes sense to treat
the off diagonal elements perturbatively.

We expand the density matrix
\begin{equation}
\rho_{ij} = D_{ij} + \epsilon_{ij}
\end{equation}
where $D_{ij}$ is the diagonal part of the density matrix and $\epsilon_{ij}$ is
the remaining off-diagonal terms, with $\epsilon_{ii}=0$. The entropy is $S(\rho) = - \tr (\rho \ln \rho))$.

We expand the entropy around $\rho = D$ to second order in $\beps$. This is calculated in Appendix \ref{app:OffDiagEntroCalc}
and gives
\begin{equation}
\label{eq:S2sum}
S_2(\rho) = \frac{1}{2} \sum_{n \ne m}  F(\rho_{nn},\rho_{mm}) \rho^2_{nm} .
\end{equation}

The function $F(x,y)$, defined in Eq. \ref{eq:DefF}, is symmetric in its arguments
and is peaked along the line $x=y$, where it has the value $F(x,x) = 1/x$. As $x$ and $y$
go to zero, the summand in Eq. \ref{eq:S2sum} is well behaved as can be checked as follows.

By Schwartz's inequality Eq. \ref{eq:rhoCoeffs} implies

\begin{equation}
\rho^2_{ij}  =  (\sum_m C_{im} C_{jm})^2  \le \rho_{ii} \rho_{jj}
\end{equation}

Therefore
\begin{equation}
|F(\rho_{nn},\rho_{mm}) Re(\epsilon^2_{nm})| \le |F(\rho_{nn},\rho_{mm}) \rho_{mm} \rho_{nn}|
\end{equation}
And it is easily seen that the function $F(x,y) x y$ is well behaved for small $x$ and $y$.

Now we are in a position to estimate the order of the off-diagonal contribution $S_2$
to the total entanglement entropy $S$ using Eq. \ref{eq:S2sum}.
With $N^2$ terms in the sum, $\rho^2_{nm}$ estimated using Eq. \ref{eq:rhoijEstimate},
 and $F$ contributing $O(\ln \rho_{ii}/\rho_{ii}) = O(\ln N/(1/N)) = O(N\ln N)$, 
this gives
\begin{equation}
S_2 \sim N^2  N \ln N (\frac{1}{N^2 M}) \sim O((\ln N) N/M)
\end{equation}

The diagonal term, which is also the thermodynamic entropy is $\ln N$. 
Therefore for $N \ll M$ the off-diagonal contribution to the entropy is negligible and the entropy
is given by the ensemble result. Also note that for a homogeneous system in order for $S_2$ to be small compared to
the canonical result, one does not require $n \ll m$. Because
$\ln(N/M)  \propto n - m$ one instead requires that $m-n$ is large. For a macroscopic system, $m-n$ can be made very large
while $(m-n)/m$ can be very small. This suggests that in the limit of large $n$, the entanglement entropy
will be equivalent to the canonical entropy for $n/m < 1$.

Because the entanglement entropy of $\phi$ is identical~\cite{BennettBernsteinPopescuSchumacher}  to that of $\theta$, for $n/m > 1$, the
entropy obtained from the entanglement entropy becomes that of system $\theta$.

\section{Discussion}

There are many definitions of th entropy, and it is often couched
in terms of the lack of information about a system. 
Classically, the relationship between information and the system's state is
straightforward.
If the microscopic state of a system
is completely characterized, this means that we have all possible information
about it. But a gas applies the same time-averaged pressure to a
piston irrespective of the experimenter's state of ignorance. The entropy
used for thermodynamic purposes is not dependent on our knowledge of the
system, so that derivatives of the entropy with respect to parameters
such as volume  via Eq. \ref{eq:SmicroDeriv}, give us experimentally
measurable quantities such as the pressure. This is why in this
work I have concentrated on understanding the {\em thermodynamic} entropy. 
This entropy is calculable through standard means, such as the canonical
ensemble.

However the situation becomes more murky when considering quantum mechanics.
The process of obtaining a precise state involves measurement, which couples
the system of interest to another system. This changes its state. Therefore
the act of measuring a system's energy precisely, so as to put it into an energy
eigenstate, might then effect its thermodynamic entropy. 

However for finite temperature systems, the above results suggest that
a generic many body system in an energy eigenstate has a well defined
thermodynamic entropy that is calculable using the usual statistical
mechanical methods, for example, through the canonical ensemble. From
the above argument, this is not obvious and it is virtually impossible to
test this experimentally on a large system because the separation between
energy levels is exponentially small in the number of degrees of freedom.

If a system starts out in some generic pure state with many different
components in different energy eigenstates, and one measures the energy
precisely, this puts it into an energy eigenstate.  One can construct an
argument suggesting that contrary to the claims here, the entropy after
such a measurement would be greatly effected.  If the system starts out
with an energy spread over some width, for example corresponding to the
wavelength of a box size, it is the sum of an exponentially large number
$N  \propto \exp(S)$ energy eigenstates $|e_i\rangle$ so that $|\psi>
= \sum_i a_i |e_i\rangle$ with some coefficients $a_i$.  If the energy
is measured to sufficient accuracy so as to put into a single
energy eigenstate, then this has reduced the number of coefficients
in this sum down to one. The energy measurement outputs a number of
great accuracy describable by a minimum of $\log_2(N)$ bits. This is
a large number that is extensive in the size of the system. This much
information being produced, and the collapse of the wave function to
a single energy eigenstate, might give one reason to believe that the
thermodynamic entropy has been reduced by an extensive amount. 

However there are two problems with this argument. First as we argued
above for the classical regime, the act of measurement process does not
alter the thermodynamic entropy.  Second, a measurement by itself does not
necessarily cost any energy. As shown by Bennett~\cite{Bennett}, it does
not cost energy to find the state of a two-state system, if this is done
sufficiently slowly. If the apparatus was originally in a known standard
state, it will find itself in a different state after the measurement
process, and that state will depend on the outcome of the measurement. In
order for further measurements to take place, the apparatus needs to be
reset which means that the phase space of the measurement apparatus must
be contracted, which will cost an energy $k_B T/2$.

One can also see that putting the system in contact with a small system
of order just a few degrees of freedom  at the same temperature will
immediately destroy the energy eigenstate, returning it to a wave function
with many energy components. Therefore, it is not plausible that measuring
the system to this accuracy could reduce its entropy by a macroscopic
amount. As far as interaction with other systems, it is expected to
behave as a more generic pure state at the same temperature.

I have calculated the entanglement entropy of an
energy eigenstate with an energy equivalent to a system
at finite temperature, to leading order in the size of the system.
For notational simplicity, I have occasionally assumed that the
system is homogeneous, however the results should apply for
non-homogeneous systems. The complete entropy is obtained
by dividing the system several ways into a larger and a smaller
subsystem and calculating the entanglement entropy between them. 

There will be corrections to this prescription that become
small for large system size.  Exact results for zero temperature
systems is an indication of the presence of sub-extensive
terms that have many interesting applications~\cite{WolfeEntanglement,SrednickiEntanglement}.
The methods used here do not easily give non-extensive corrections and
consequently do not give these zero temperature results.

The result that the entanglement entropy of a finite temperature system is
equal to the thermodynamic entropy can be shown rigorously to be the case
in special cases~\cite{CalabreseCardyReview,CalabreseCardyQFT,Korepin}. This
was shown to be the case for 1+1 dimensional conformal field theories in a finite
temperature mixed state ensemble, where the system of interest is
connected to an infinite system. It is not surprising that in this limit,
one obtains the usual entropy as a canonical ensemble has been used which
acts, in effect, as a heat bath. In this paper, we analyzed the case where
the system is in an energy eigenstate and reached the same conclusion
concerning the entanglement entropy. In this case, there is no heat
bath, only the coupling of the subsystem of interest to the rest of the
system. It is not obvious that such a coupling should be enough to result
in the answer one obtains for a mixed state.  The model that we used was
similar to that of previous work~\cite{DeutschQuantStatMech}, and assumes
that the coupling between the two parts of the system $\phi$ and $\theta$
is sufficiently chaotic to be describable by random matrices. This is certainly an
approximation and for short range forces, the interactions between
the two should be taking place only on the interface between the two subsystems. 
However I conjecture that the result is robust enough to apply so such
cases. Numerical work should in principle, be able to test the arguments
presented here.

We also looked at the effects of external forces that are slowly applied
to an energy eigenstate.  We argued that such states are no different
thermodynamically than for generic pure states. A many body energy eigenstate responds
to external perturbations through the flow of heat obeying the usual
relationship between the change in heat and the change in entropy.

If it turns out that the entanglement entropy is not the correct measure
of the thermodynamic entropy for energy eigenstates, the arguments presented
in Sec.\ref{sec:WaveFunctionThermo} make it likely that there is another
prescription involving solely the wave function that should determine the
entropy.

\appendix

\section{Range of Potential Matrix Elements}
\label{app:RangeOfPotential}

In this appendix we justify in more detail the cutoff on the random matrix
used in calculations. We are interested in determining how $\langle E | V | E'\rangle$
varies with increasing $E - E' $. The quantity we wish to compute is
$\langle\langle E | V | E' \rangle\rangle_{E , E'}$
Here the second set of brackets denotes a 
microcanonical average over both $E  $and $E '$.
$V$ is taken to be of the form (3),  and we consider a system of 
identical particles that are either fermions or bosons. 
Label the eigenstates of a single particle by $i$. The energy in that 
state is labeled $e_i$, the total number of particles in state 
$i$ is $n_i$ and $n_i \le 1$ for the case of fermions. 
In second quantized notation the total wave-function can be written as 
\begin{equation}
|E> = \prod_i \frac{{a^\dagger_i }^{n_i}}{\sqrt{n!}} |0>
\end{equation}
where $|0>$ is the ground state. The potential $V$ can also be written in 
second quantized form as 
\begin{equation}
V =  \sum_{j,k,l,m}  V_{jklm} a^\dagger_j a^\dagger_k a_l a_m
\end{equation}
where
\begin{equation}
V_{jklm}  =  \int \psi^\star_j (r) \psi^\star_k (r') V(r-r') \psi_l (r) \psi_m (r') dr dr'
\end{equation}
and in this case, $\psi_j$ denotes a plane-wave with wave vector indexed by $j$.
Thus the quantity we wish to compute is
\begin{equation}
\langle\langle E | V | E' \rangle\rangle_{E , E'} =  
\frac{1}{ n(E)~n(E')} \sum_{n_i's}  \sum_{{{n'}_j}'s}
\delta ( E - \sum_i n_i e_i ) \delta (E' - \sum_i n'_i e'_i ) \langle E | V | E' \rangle
\end{equation}
where $n(E)$ is the appropriate normalization. In the above, 
$\sum_{n_i's}$ means the sum over all possible combinations of
$n_i's$ with the constraint $\sum_i n_i = n$.
Writing the above equation 
in second quantized form and taking the inner products gives
\begin{eqnarray}
\langle\langle E | V | E' \rangle\rangle_{E , E'}   = & \nonumber \\
\frac{1}{ n(E) n(E')} & \sum_{j,k,l,m}  \sum_{n'_l , n'_m}
  \sum_{n_i's} \delta ( E - \sum_i n_i e_i ) \sqrt{n_j n_k n'_l n'_m} \delta (E-E'-(e_j + e_k - e_l -e_m )) V_{jklm}\nonumber \\
= \frac{1}{n(E')}  & \sum_{j,k,l,m}  \sum_{n'_l , n'_m} \langle \delta (E-E'-(e_j + e_k - e_l -e_m )) \sqrt{n_j n_k n'_l n'_m} V_{jklm} \rangle_E
\end{eqnarray}
where the last bracket denotes a microcanonical average at energy $E$.
As long as $|E-E'| \ll E$, and $n$ is large,
the microcanonical average here can be replaced
by a canonical average at the appropriate temperature T as given in (2.4)
where 
\begin{equation}
S~=~ \ln ( \sum_{n_i's}  \delta (E- \sum_i n_i e_i ) 
\end{equation}
If we consider potentials $V(r)$ which have a Fourier transform that is
bounded, then so is $V_{jklm}$ as the single particle eigenstates are
plane waves. We can therefore bound the above equation by
\begin{equation}
\frac{1}{n(E')} \sum_{j,k,l,m} \sum_{n'_l , n'_m} \langle \delta (E-E'-(e_j + e_k - e_l -e_m )) \sqrt{n_j n_k n'_l n'_m} \rangle_T
\end{equation}
Now consider what happens for $E-E' \gg T$. In this limit 
it is straightforward to
substitute in the appropriate Bose or Fermi distributions for
each $n_i$ and perform the summations,
but the asymptotic result can be seen by the following argument.
The above average only has contributions to it when 
$e_j + e_k = e_l + e_m + E -E'$. If $E'$ is kept fixed
and E is increased then the minimum energy needed to obtain an contribution
occurs when $e_l = e_m = 0$, so that
$e_j + e_k =  E -E'$.(Here we are setting the ground state 
energies equal to zero.) As E is increased the weight of having
such a configuration is given by the appropriate Bose or Fermi distributions
which asymptotically give a weight of $ \exp (-e_j - e_k ) =
\exp ( -(E-E')/T) $. Considering larger $e_l$ and $e_m$ does not
change the above exponential dependence, but just the overall pre-factor.

\section{Relation Between Entropy and Forces}
\label{app:RelEntrForce}

Here we present a derivation of Eq. \ref{eq:SmicroDeriv} that is shorter than
other treatments that the author is aware of. 

Consider the integral of the density of states
\begin{equation}
I(E) = \int_{-\infty}^E e^{S(E')} dE' = \int_{-\infty}^E  \tr \delta(E-H) = \tr \theta(E-H)
\end{equation}
Where the last equality uses the Heaviside function $\theta$.

Because the entropy is rapidly increasing for a system with a large
number of degrees of freedom $n$, we expand it about $E$, in the exponent
of the above integrand, 
\begin{equation}
  S(E')  = S(E) + \frac{\partial S}{\partial E}  (E'-E) + \dots
   =  S(E) + \frac{1}{T}  (E'-E) + \dots
\end{equation}
So that for large $n$, 
\begin{equation}
I(E) = \int_{-\infty}^E e^{S(E')} dE' = T e^S(E)
\end{equation}

Now consider a Hamiltonian that depends on a parameter $x$,
\begin{equation}
\frac{\partial \ln I(E)}{\partial x} = - \frac{\tr (\delta(E-H) \frac{\partial H}{\partial x})}{T e^S}
=  - \frac{1}{T} \langle\frac{\partial H}{\partial x}\rangle
\end{equation}
One can relate the left hand side to the entropy as follows
\begin{equation}
\frac{\partial \ln I(E)}{\partial x} = \frac{\partial S(E)}{\partial x}  +  \frac{\partial T}{\partial x}    
\end{equation}
which for large $n$ becomes $\partial S/\partial x$, giving Eq. \ref{eq:SmicroDeriv}.

\section{Calculation of Off Diagonal Component of Entropy}
\label{app:OffDiagEntroCalc}

We wish to calculate the effects of small off diagonal elements of the density matrix
on the entropy $S(\rho) = - \tr (\rho \ln \rho))$, by writing 
$ \rho_{ij} = D_{ij} + \epsilon_{ij} $.

To simplify the expansion, first define
$f(x) = -x \ln(x)$. There is no power series expansion of this about  $x=0$, but we
can regularize it to allow such an expansion, for example $f_r(x) = f(x+\delta)$. We
will see that at the end, we can take the limit $\delta \rightarrow 0$ without
difficulty. With a regularized $f$, we can it expand it as
\begin{equation}
f(x) = \sum_{n=0}^\infty a_n x^n
\end{equation}

\begin{equation}
S(\rho) = \tr (f_r(\rho)) = \tr (f_r(\bD + \beps)) = \sum_{n=0}^\infty a_n (\bD + \beps)^n
\end{equation}

The $\epsilon^0$ term yields $S_0 = \tr(f(\bD))$, as expected. 
The $\epsilon^1$ term is zero because $\tr(\bD^n \beps) = \sum_i D^n_{ii} \epsilon_{ii} = 0$.

To obtain the $\epsilon^2$ contribution, we note that $\beps$ and $\bD$ are in general
non-commuting, and therefore we must preserve matrix ordering when expanding $(\bD+\beps)^n$.
Denoting terms second order in $\beps$ by $S_2$, we have

\begin{equation}
S_2(\rho) = \tr \sum_{n=0}^\infty a_n \sum_{i,j,k=0}^\infty \bD^{i} \beps \bD^j \beps \bD^k \delta_{i+j+k,n-2}
\end{equation}

However because for two matrices $A$ and $B$, $\tr(AB) = \tr(BA)$, we can reorder
a given term so that final matrix is always $\beps$ , so that the summand
becomes $\bD^{i+k}\beps  \bD^j \beps$. This reordering generates $i+k+1$ such terms. So simplifying
the indices gives
\begin{equation}
\label{eq:s2trsum}
S_2(\rho) = \tr \sum_{i,j=0}^\infty (i+1) a_{i+j+2} \bD^{i} \beps \bD^j \beps
\end{equation}
We now commute the trace with the summations and perform it first. It has the
general form $\tr(A\beps B\beps)$ with $A$ and $B$ diagonal and real. In this case this
can easily seen to be 
\begin{equation}
\sum_{n,m} A_{nn} B_{mm} \epsilon^2_{nm}  
\end{equation}
Switching indices $m$ and $n$ and using the fact that in the situation discussed
here, $\beps$ is hermitian and $A_{nn} B_{mm} = A_{mm} B_{nn}$ yields 
\begin{equation}
\tr(A\beps B\beps)  = \sum_{n,m} A_{nn} B_{mm} e^2_{nm}  
\end{equation}
where $e^2_{nm} = Re (\epsilon^2_{nm})$. Using this in Eq. \ref{eq:s2trsum}
gives
\begin{equation}
S_2(\rho) = \sum_{n,m=0}^\infty \left(\sum_{i,j=0}^\infty (i+1) a_{i+j+2} D^i_{nn}  D^j_{mm}\right) e^2_{nm} .
\end{equation}
Because $\bf e$ is symmetric, the indices $m$ and $n$ can be exchanged, and so
after switching the dummy variables $i$ and $j$ we can add this relabeled
expression to the above, giving 
\begin{equation}
S_2(\rho) = \frac{1}{2} \sum_{n,m=0}^\infty \left( \sum_{i,j=0}^\infty (i+j+2) a_{i+j+2} D^i_{nn}  D^j_{mm}\right) e^2_{nm} .
\end{equation}
We can write the term in parentheses as $F(D_{nn},D_{mm})$ where
\begin{equation}
F(x,y) = \sum_{i,j=0}^\infty (i+j+2) a_{i+j+2} x^i y^j
\end{equation}
and we wish to find a closed form expression for this in terms of $f_r(x)$ which
have the $a_n$'s coefficients as it power series expansion.  To do this, we
group all terms according the value of  $n \equiv i+j$.
Therefore
\begin{equation}
\label{eq:Fxysum}
F(x,y) = \sum_{n=0}^\infty (n+2) a_{n+2} (x^n + x^{n-1}y + \dots xy^{n-1} + y^n)
   = \sum_{n=0}^\infty (n+2) a_{n+2} \frac{(x^{n+1}-y^{n+1})}{x-y}
\end{equation}
This can be further simplified by noting that
\begin{equation}
   \sum_{n=0}^\infty (n+2) a_{n+2} x^{n+1} = 
   \sum_{l=2}^\infty l a_{l}  x^{l-1} =  f_r'(x) - f_r'(0)
\end{equation}
where the prime denotes the first derivative. Substituting this into Eq. \ref{eq:Fxysum}
gives 
\begin{equation}
F(x,y) = \frac{f_r'(x)-f_r'(y)}{x-y}
\end{equation}
Note that because of the cancellation of $f_r'(0)$ we can now take the limit as 
$\delta \rightarrow 0$ and use $f = -x \ln x$ instead. This yields
\begin{equation}
\label{eq:DefF}
F(x,y) = \frac{\ln x- \ln y}{x-y}
\end{equation}
Summarizing, we have shown that the second order correction to the entropy due to
off-diagonal components of the density matrix $\beps$ is
\begin{equation}
\label{eq:AppendixS2sum}
S_2(\rho) = \frac{1}{2} \sum_{n,m=0}^\infty  F(\rho_{nn},\rho_{mm}) Re(\epsilon^2_{nm}) .
\end{equation}

\end{document}